\documentclass[preprint]{elsarticle}
\usepackage{url}
\usepackage{graphicx}            
\usepackage{float}
\begin{document}
\begin{frontmatter}
\title{Generating and solving the mean field and pair approximation equations
in epidemiological models}
\author[nrc,uw]{Murray E.~Alexander}
\ead{Murray.Alexander@nrc-cnrc.gc.ca}
\author[uw,witp]{Randy Kobes}
\ead{r.kobes@uwinnipeg.ca}
\address[nrc]{Institute for Biodiagnostics, National Research Council of Canada,
Winnipeg, Manitoba, R3B 1Y6, Canada}
\address[uw]{Physics Department, University of Winnipeg, Winnipeg, Manitoba, R3B 2E9, Canada}
\address[witp]{Winnipeg Institute for Theoretical Physics, 
University of Winnipeg, Winnipeg, Manitoba, R3B 2E9, Canada}
\begin{abstract}
The pair approximation is a simple, low--order method to incorporate effects of local spatial
structure in epidemiological models. However, since for $K$ state variables in a model there
are $K(K+1)/2$ equations in the pair approximation, generating these equations, although 
straightforward, can become tedious. In this paper we describe two programs written in Perl to
simplify this process -- one
to construct the equations, and the other to generate Matlab/Octave functions to
numerically integrate the equations using a positivity-preserving method. 
A third utility program is also included which generates a Maple
file that can be used within the Maple symbolic manipulation program to simplify algebraically
some of the terms in the generated file describing the equations, as well as to check that
the usual combination of equations sum up to zero, which is expected in cases where the total 
population sizes are conserved.
\end{abstract}
\end{frontmatter}
\section{Introduction}
\label{intro}
Because of the many applications in technological, biological, and social
systems, interest in networks has grown significantly in recent years 
\cite{watts, strogatz, albert, newman}.
In approaches that use differential equations to study network evolution, much work 
has gone into incorporating local spatial correlations
to extend the mean field approximation. One of the simplest
methods to do this is the pair approximation \cite{rand, keeling1, keeling2,vanbaalen}.
\par
To set the framework, consider a network of $N$ nodes with links between nodes
characterized by an adjacency matrix $G_{ij}$, where $G_{ij}=1$ if nodes $i$ and $j$ are connected
and $G_{ij}=0$ otherwise. We assume bidirectional links, so $G_{ij}=G_{ji}$, and there is
no self--contact, so $G_{ii}=0$. The number of connected pairs and triplets in the network is
\begin{eqnarray}
& &{\rm number\ of\ pairs} = || G || \equiv \sum_{i, j=1}^N G_{ij} = nN \nonumber\\
& &{\rm number\ of\ triplets} = || G^2 || -  {\rm Tr\ } (G^2)
\end{eqnarray}
where $n$ is the average number of neighbours per node. The triplets, comprising of nodes
connected by two links, will include as a subset triangles, which are three linked nodes
with the same start and end. A parameter $\phi$ is introduced to characterize the
ratio of triangles to triplets:
\begin{equation}
\phi = \frac{ {\rm number\ of\ triangles}} { {\rm number\ of\ triplets} }
= \frac{ {\rm Tr\ } (G^3) } { || G^2 || - {\rm Tr\ } (G^2) }
\end{equation}
With $0 \le \phi \le 1$, networks with values of $\phi$ close to 1 are highly clustered, while
low values of $\phi$ close to 0 show little such structure. The value of $\phi$,
together with $n$, the average number of
neighbours per node, will be considered fixed parameters for a given network, and
will give an indication of the amount of spatial structure present.
\par
Models describing the evolution of networks that incorporate such spatial correlations
can be formulated as follows. Suppose there are a set of variables associated with
each node, generically
labelled by $A$, which is 1 if the node is of type ``$A$''. Introduce singlets, doublets, and
triplets as follows:
\begin{eqnarray}
& &{\rm singlets\ of\ type\ A} = [A] = \sum_i A_i \nonumber\\
& &{\rm doublets\ of\ type\ AB} = [AB] = \sum_{i, j} A_i B_j G_{ij}  = [BA] \nonumber\\
& &{\rm triplets\ of\ type\ ABC} = [ABC] = \sum_{i, j, k} A_i B_j C_k G_{ij} G_{jk}
=[CBA]
\label{rels}\end{eqnarray}
One can show
\begin{eqnarray}
& &\sum_A [A] = N \nonumber\\
& &\sum_B [AB] = n[A] \nonumber\\
& &\sum_C [ABC] = \frac{n(n-1)}{N} [A][B]
\label{sum}\end{eqnarray}
Consider now, for example, the $SIR$ model, in which a given node can exist in one
of three states: susceptible ($S$), infected ($I$), or recovered ($R$). A susceptible
node will become infected, with rate characterized by $\beta$, while infected nodes
recover with a rate characterized by $\gamma$. Equations describing these
transitions are
\begin{eqnarray}
& &\frac{dS}{dt} = -\beta SI \nonumber\\
& &\frac{dI}{dt} = \beta SI -\gamma I\nonumber\\
& &\frac{dR}{dt} = \gamma I 
\label{sirmf}\end{eqnarray}
Note that the sum of these equations vanishes, which reflects the fact that
the total population remains constant. These equations describe the
evolution of the appropriate singlets $[S]$, $[I]$, and $[R]$, and can be
used to derive the corresponding equations for the doublets \cite{keeling2}:
\begin{eqnarray}
& &\frac{d\, [SS]}{dt} = - 2\tau [SSI] \nonumber\\
& &\frac{d\, [SI]}{dt} =  \tau [SSI]  -\tau[ISI] -\tau[SI] - \gamma [SI]\nonumber\\
& &\frac{d\, [SR]}{dt} = \gamma [SI] -\tau [RSI] \nonumber\\
& &\frac{d\, [II]}{dt} = 2\tau [ISI] + 2\tau[SI] - 2\gamma [II]\nonumber\\
& &\frac{d\, [IR]}{dt} =  \tau [RSI] +\gamma [II] - \gamma [IR]\nonumber\\
& &\frac{d\, [RR]}{dt} = 2\gamma [IR]
\label{sirpa}\end{eqnarray}
where $\tau = \beta / n$. Since triplets appear in this equation, we require
a closure approximation to express the triplets in terms of doublets; a
common one used in this regard is parameterized by the number of
nearest neighbours $n$ and the ratio $\phi$ of triangles to triplets \cite{keeling2}
\begin{equation}
[ABC] \approx \zeta \frac{ [AB] [BC]} { [B]} \left[ (1-\phi) + 
\phi \frac{N}{n} \frac{ [AC]} { [A] [C]} \right]
\label{keeling}\end{equation}
where $\zeta =(n-1)/n$.
\par
Although the procedure to generate the equations in the pair approximation is 
straightforward \cite{keeling2}, it can become tedious; if there are $K$ mean field
equations, there will be $K(K+1)/2$ equations in the pair approximation. In the next section
we describe the use of programs, written in Perl, to assist in these tasks: one program to
generate the equations, and another to construct Matlab/Octave functions that can
be used to numerically 
solve the equations. Additionally, a utility Perl script is supplied which, 
based on the file used in the other programs to describe the equations, will generate
a text file that can be loaded into the Maple symbolic manipulation program to provide
algebraic simplifications of the terms in the equations, as well as check that the
standard combination of terms in the equations sums up to zero, which is 
expected when the total population size is constant (as in the $SIR$ model described
above).
\section{Programs}
After explaining how to install the programs, we will describe their use
in the order they would normally be used.
\subsection{Installation}
The programs are available at \url{http://physics.uwinnipeg.ca/rkobes/} in a
zipped archive called {\tt pa\_programs.zip}. Unzipping the archive will create
a subdirectory {\tt pa\_programs} containing a number of files:
\begin{itemize}
\item Three Perl scripts -- {\tt make\_eqns.pl},  {\tt make\_mfile.pl}, and
{\tt maple\_chk.pl} -- described in this paper;
\item Documentation in {\it HTML} format for all three scripts;
\item Sample {\it JSON} configuration files -- {\tt sir.json}, {\tt siri.json}, and {\tt dr.json} --
for, respectively, the $SIR$, $SIRI$, and a drug resistant model discussed in this paper.
\item A pdf copy of this paper
\item a {\tt README} file summarizing the installation procedure
\end{itemize}
Information on the usage
of the programs can be obtained by the {\tt perldoc} command in Perl: in a
shell command window, type 
{\tt perldoc make\_eqns.pl}, for example. HTML versions of the
documents, suitable for viewing within a web browser, can be made with the
{\tt pod2html} utility: {\tt pod2html --infile make\_eqns.pl --outfile make\_eqns.html},
and similarly for the other scripts.
\par
As Perl is very good at text manipulations, and has fairly advanced regular expression
support, it is normally available on most modern Unix--based systems, including Linux
and MacOS (with the developer tools installed). The most popular binary Perl
distribution is that distributed by ActiveState: \url{http://www.activestate.com/activeperl};
this includes a Windows version. The programs should run on versions of Perl
back to 5.6. The only requirement beyond the core Perl installation is a
Perl module called {\tt JSON}. On systems using the ActivePerl binary distribution,
{\tt JSON} can be installed with the {\tt ppm} utility: in a shell command
window, type {\tt ppm install JSON}. For especially Linux--based systems with a 
package manager such as {\tt rpm} or {\tt apt-get}, there may be pre--built binary
distributions available. On other systems the module will have to be built and
installed from the sources, which are available at
\url{http://search.cpan.org/dist/JSON/}. How to install modules from sources
 is described at either
\url{http://perldoc.perl.org/perlmodinstall.html} or
\url{http://www.perlmonks.org/index.pl?node_id=128077}. This latter link
also describes what to do in cases where a user doesn't have permission
to perform a system--wide module installation.
\par
One aspect of the running of the Perl scripts is worth noting. As it stands, the
scripts, containing a {\tt .pl} extension, can be run from a shell command window
 as, for example,
{\tt perl make\_eqns.pl}. This can be simplified as follows. On a Unix--based system,
if one makes a line such as {\tt \#!/usr/bin/perl} the first line of a Perl script (where
{\tt /usr/bin/perl} is the Perl interpreter, including the full path), 
remove the {\tt .pl} extension from the filename, make the script executable
by running, for example, {\tt chmod u+x make\_eqns}, and then place this file
in a directory appearing in your {\tt PATH} environment variable. The script can
be run simply as {\tt make\_eqns}. On a Windows--based system, there is a Perl
utility {\tt pl2bat} which, if run as, for example, {\tt pl2bat make\_eqns.pl}, will create
a DOS batch file {\tt make\_eqns.bat}. Placing this file on a directory appearing in
your PATH environment variable will also enable the script to be run simply as
{\tt make\_eqns}.
\subsection{The {\tt make\_eqns.pl} program}
This program starts from a user--supplied input file, written in {\it JSON},
that specifies the system under consideration, and outputs {\it JSON} files describing the
mean field and pair approximation equations. We begin by describing
the structure of this input.
\subsubsection{The {\it JSON} file}
 {\it JSON} (JavaScript Object Notation)  is a lightweight text-based 
open standard designed for human-readable data interchange.
 {\it JSON} files are ordinary text files, which by convention normally
 have a {\tt .json} extension.
As such, they can be created with any text editor; if one uses a word processor such
as Microsoft Word${}^\copyright$, one must ensure to save the file in text mode, with a
final {\tt .json} extension. 
As an overview, there are two basic structures in {\it JSON}:
\begin{itemize}
\item
A container, enclosed within opening and
closing curly braces, consisting of a collection of key/value pairs. 
A key and its associated value are separated by a colon,
and the key/value pairs are separated by commas.
\item
An ordered list of items, which is enclosed within opening and closing square
brackets. Individual items are separated by commas.
\end{itemize}
The basic data types are
\begin{itemize}
\item
A number, which can be an integer or real.
\item
A string, which is a double-quoted group of Unicode characters
with backslash escaping.
\item
A Boolean value, which can be {\it true} or {\it false}.
\item
The {\it null} value.
\end{itemize}
See \url{http://www.json.org/} for a general discussion and further
links. 
 \subsubsection{$SIR$ model}
\begin{figure}[H]
\begin{center}
\includegraphics[width=4in]{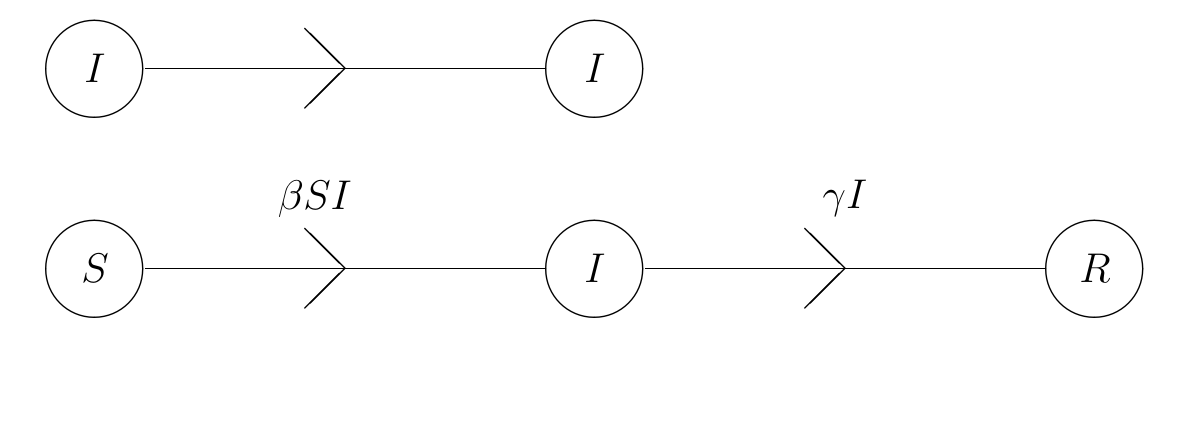}
\caption{The basic $SIR$ model.}
\label{sir}
\end{center}
\end{figure}

 The basic $SIR$ model is described by two transitions: an $S \to I$, at rate $\beta SI$, and
 an $I\to R$, at rate $\gamma I$. The $S\to I$ transition needs a ``spectator'' $I$ node
 to proceed (susceptible nodes become infected only when they come into contact with
 an infected node), but the $I\to R$ transition proceeds without a ``spectator'' (infected nodes
 can recover on their own). Fig.~\ref{sir} illustrates this model.
The corresponding {\it JSON} file for this model appears below.
\begin{verbatim}
##############################################
# file sir.json, specifying transitions in the SIR model
{
  "S": {  
     "target": "I",
     "link": "beta",
     "needs": "I",
 },
  "I": { 
     "target": "R",
     "link": "gamma",
 },
  "pa_parameters": {
     "beta": "tau",
 },
  "first" : "S",
}
##############################################
\end{verbatim}
Note that new lines and leading and trailing white space on a line
have no significance in the {\it JSON} file; they are only present here for ease
of human readability.
\par
Before discussing the structure of this file, an important
point should first be noted. The above file differs in two aspects
from the strict {\it JSON} specification:
\begin{itemize}
\item
End-commas in list items can optionally appear.
\item
Shell-style comments beginning with a {\tt \#} can be used.
\end{itemize}
This file would be ruled invalid if processed by a parser following the strict
{\it JSON} specification. However, the {\tt JSON} Perl module used in this script
has a {\it relaxed} mode that is enabled that allows for these two non-standard
extensions. The {\it JSON} output files from this script follow the
strict {\it JSON} specification without these non-standard extensions.
\par
The basic features illustrated in this example
are as follows.  
\begin{itemize}
\item
Except for the special {\tt "pa\_parameters"} and {\tt "first"} keys, a key such as, 
for example, {\tt "S":}, indicates the start of
the description for the {\tt S} node.
\item
The properties of each node are given (in this example)
as a container consisting of key/value pairs.
There are three possible values of the keys:
\begin{itemize}
\item
The {\tt "target"} field, which is required, indicates 
the target of the transition.
\item
The {\tt link} field, which is required, indicates the parameter characterizing
the transition. The associated value can be a single variable
or a mathematical expression
following the rules of Matlab/Octave -- for example, {\tt "2*g*(1-p)"}, expressing 
combinations of several variables.
\item
The {\tt "needs"} field, which is optional, indicates whether or not another state
variable is needed to cause the transition.
\end{itemize}
\item
The purpose of the {\tt "first"} key arises because key/value pairs
in a container will not be sorted. If a {\tt "first"} key is present
in the {\it JSON} file, this will be transferred over to a {\tt "first"} key
in the output files. The presence of such a key in these files, when
processed by {\tt make\_file.pl}, will cause the singlets in the Matlab/Octave
update functions to be ordered such that the value of the {\tt "first"} key
appears first, after which the rest are sorted alphabetically. If a
{\tt "first"} key is not present, the singlets will be sorted alphabetically.
\item
The purpose of the {\tt "pa\_parameters"} key is as follows.
Often, when deriving the equations in the pair approximation, a
parameter arising in the mean field approximation gets replaced by
another parameter in the pair approximation, where the two are related
through the number of nearest neighbors, {\tt n}. For example, in the
SIR model, the {\tt beta} parameter in the mean field case gets replaces
by {\tt tau = beta / n} in the pair approximation. In the json file,
one can list parameters related in this way through a container such as
\begin{verbatim}
  "pa_parameters": {
      "beta": "tau",
  },
\end{verbatim}
When constructing the equations in the pair approximation, the
parameter {\tt "beta"} will be replaced by {\tt "tau"}. The relationship
between {\tt "beta"} and {\tt "tau"} must be specified by the user in the
main program.
\end{itemize}
\subsubsection{$SIRI$ model}
\begin{figure}[H]
\begin{center}
\includegraphics[width=4in]{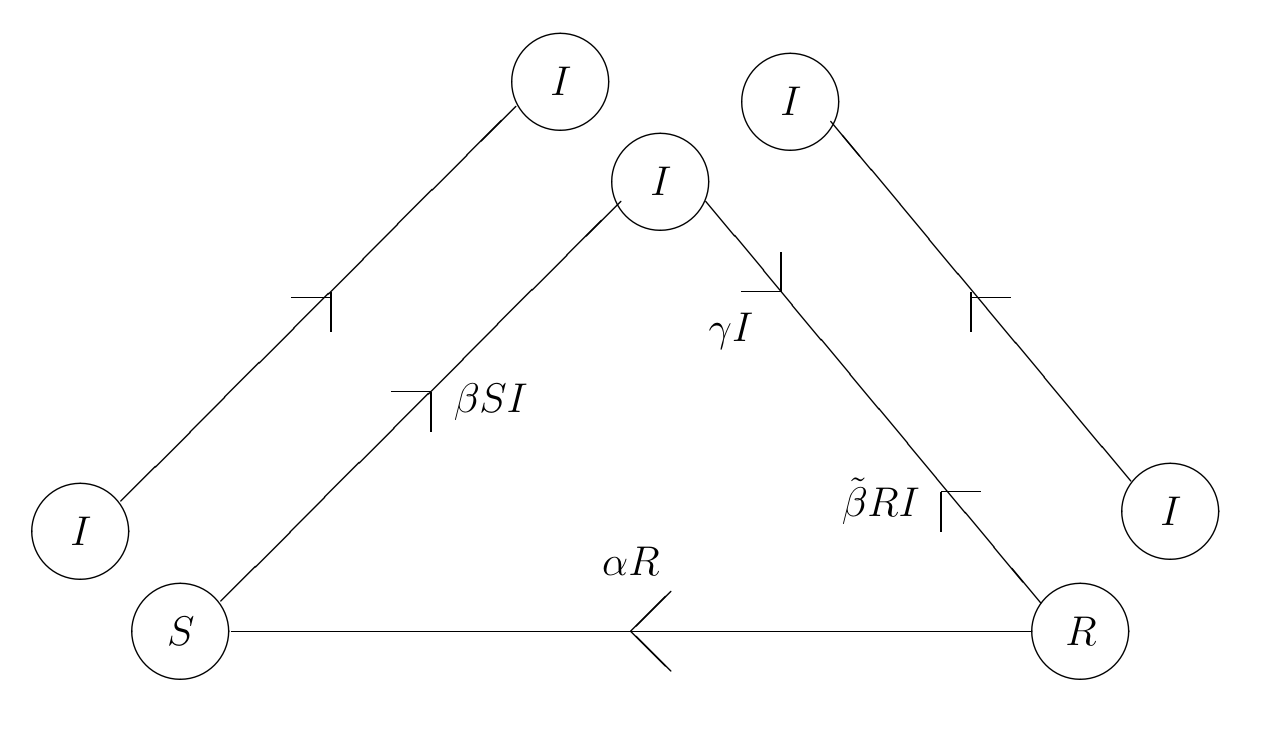}
\caption{The basic $SIRI$ model.}
\label{siri}
\end{center}
\end{figure}

The $SIRI$ model extends the $SIR$ model by having, as well as the transitions
$S \to I$, at rate $\beta SI$, and $I\to R$, at rate $\gamma I$, of the $SIR$ model, two
additional transitions: $R\to S$, at rate $\alpha R$, representing a recovered node
becoming susceptible, and $R\to I$, which proceeds in the presence of an $I$ node
at rate ${\tilde \beta} RI$, representing a recovered node becoming infected again.
 Fig.~\ref{siri} illustrates this model.
The {\it JSON} file representing this model appears below.
\begin{verbatim}
##############################################
# file siri.json, specifying transitions in the SIRI model
 {
  "S": {
     "target": "I",
     "link": "beta",
     "needs": "I",
  },
  "I": {
     "target": "R",
     "link": "gamma"
   },
   "R": [
     {
       "target": "S",
       "link": "alpha",
     },
     {
       "target": "I",
       "link": "beta_tilde",
       "needs": "I",
     }
   ],
   "pa_parameters": {
      "beta": "tau",
      "beta_tilde": "tau_tilde"
   },
   "first" : "S",
 }
##############################################
\end{verbatim}
The one new feature illustrated here that is not present in the $SIR$ model is the
presence of more than one possible transition for a given node (the R node).
In such cases, the value of the {\tt "R"}
node is a list of containers, each container following
the rules for a single transition.
\subsubsection{Drug resistant model}
\label{drsection}
\begin{figure}[H]
\begin{center}
\includegraphics[width=3.5in]{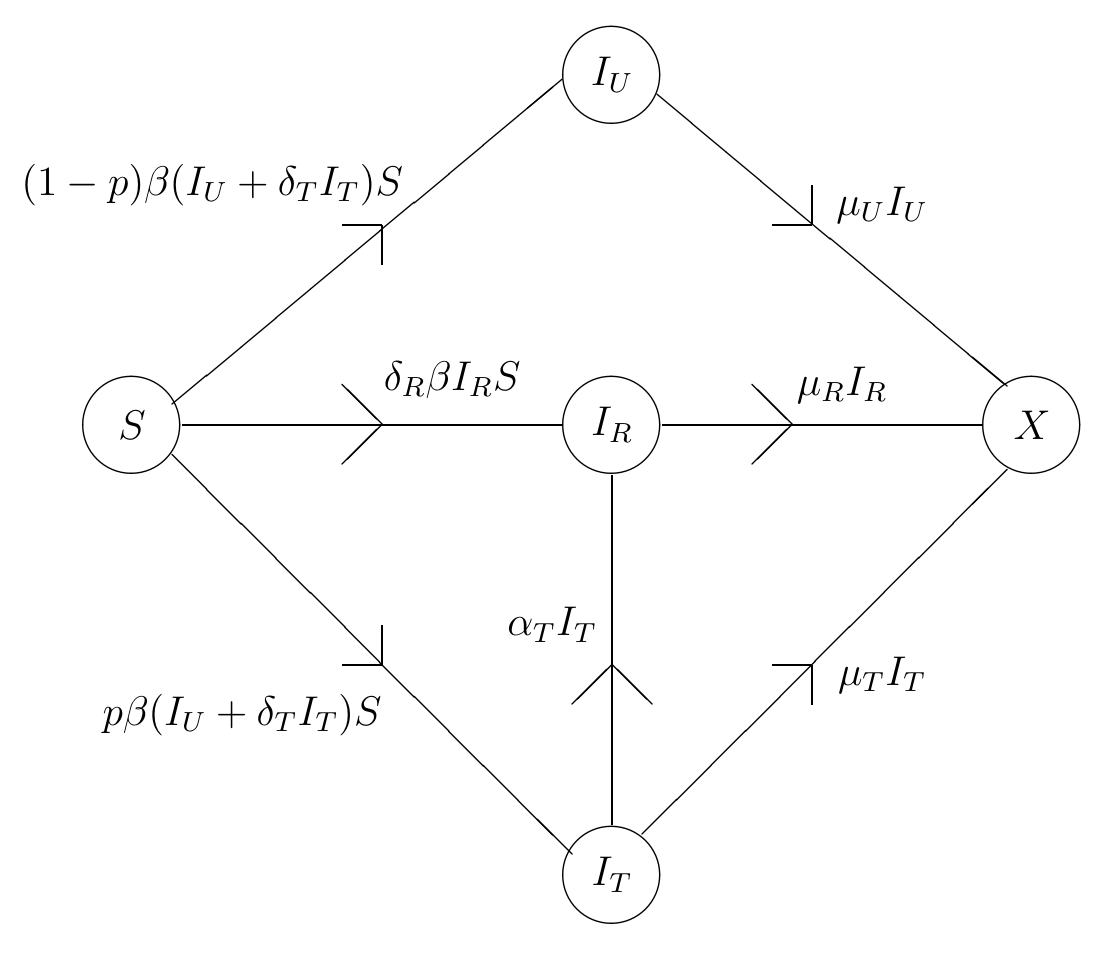}
\caption{A drug resistant model.}
\label{dr}
\end{center}
\end{figure}

There are more complicated models that involve transition rates between nodes
which are not covered by the previous two examples. For example,
in the drug resistant model considered in Ref.~\cite{murray}, there are two strains
of a pathogen, one of which is sensitive to a drug treatment, and the other is
resistant. There are subsequently three classes of infected individuals: $I_U$ and
$I_T$, representing, respectively, untreated and treated nodes infected with the
sensitive strain, and $I_R$, representing a node infected with the resistant strain.
A parameter $p$ is introduced, representing the fraction of infected individuals
who receive treatment for the sensitive strain (ie, the treatment level). The susceptible
to infected transitions therefore consist of terms:
\begin{itemize}
\item $S\to I_U$, at rate $(1-p)\beta(I_U + \delta_T I_T) S$
\item $S \to I_T$, at rate $p \beta(I_U + \delta_T I_T) S$
\item $S \to I_R$, at rate $\beta \delta_R I_R S$
\end{itemize}
where $\delta_T$ is the relative infectiousness of treated individuals 
infected with the sensitive strain, and $\delta_R$ 
represents the relative transmission fitness of the resistant strain. For simplicity,
death and recovered classes are grouped into an $X$ node, with $\mu_U$, $\mu_T$, and
$\mu_R$ parameterizing transitions from, respectively, the $I_U$, $I_T$, and $I_R$ nodes
to the $X$ node. Finally, a transition $I_T\to I_R$, characterized by rate $\alpha_T$,  is
included. This model is represented in Fig.~\ref{dr} (without the ``spectator nodes'' required in
the transitions involving $S$).
The full {\it JSON} file for this model appears in the Appendix; the 
portion describing the $S$ transitions is as follows:
\begin{verbatim}
##############################################
   "S": [
      {
         "target": "I_U",
         "link": { 
            "I_U": "(1-p)*beta", 
            "I_T": "(1-p)*beta*delta_T",
         },
         "needs": "I_U"
      },
      {
         "target": "I_T",
         "link": {
            "I_U": "p*beta",
            "I_T": "p*beta*delta_T",
         },
         "needs": "I_T",
      },
      {
         "target": "I_R",
         "link": "beta*delta_R",
         "needs": "I_R"
      },
   ]
 ##############################################
\end{verbatim}
In this case, the more complicated links specifying the
transitions to $I_U$ and $I_T$
are specified by a container, rather than as a simple scalar value
as in the previous examples. Each key/value pair within the
container is of the form {\tt "node": "parameter\_value"}.
\subsubsection{Running {\tt make\_eqns.pl}}
The {\it JSON} file described in the previous sections is then used as input
to the Perl script {\tt make\_eqns.pl}. The script may be used with
various options as follows, assuming
an input {\it JSON} file {\tt sir.json}:
\begin{itemize}
\item
{\tt perl make\_eqns.pl sir.json}: 
This will generate two text files: {\tt sir\_mf.txt}, describing the mean field equations, and {\tt sir\_pa.txt}, 
describing the pair approximation equations.
\item
{\tt perl make\_equns.pl --input sir.json --mf mf.txt --pa pa.txt}:
This does the same as the previous usage, but allows you to specify the output file names.
\item
{\tt perl make\_eqns.pl --nmax 35 sir.json}
The {\it nmax} option, which is optional and has a default of 40, specifies the maximum 
length of the lines appearing in {\tt sir\_mf.txt} and {\tt sir\_pa.txt}
that specify the equations.
\item
{\tt perl make\_eqns.pl --mfile sir.json}
The {\tt --mfile} option, if given, will run {\tt make\_mfile.pl} after
finishing, which will generate the Matlab/Octave functions that can be used to solve the
system of equations. The input to {\tt make\_mfile.pl} will be the two output {\it JSON} files
from this script. The output mfiles from {\tt make\_mfile.pl}
will have the default names. Details of generating these Matlab/Octave files,
and how they can be used, will be given in the next section.
\item
{\tt perl make\_eqns.pl --help}: This prints a brief summary of usage of the script and exits.
\end{itemize}
\subsection{The {\tt make\_mfile.pl} program}
This program starts from a user--supplied input file, written in {\it JSON},
that specifies the equations of the system under consideration, and outputs Matlab/Octave
functions that can be used to solve the equations. These input files are the output
from {\tt make\_eqns.pl} described in the previous section, although they can be originate
from another source. 
\subsubsection{{\it JSON} files}
The {\it JSON} file used here describes a system of first--order differential equations,
involving $N_X$ state variables $X_i$, with $i=1, 2, \ldots, N_X$. For the mean field model,
$X_i$ would correspond to singlets, while for the pair approximation equations, the $X_i$
 would be doublets. The equations are assumed to have the form
\begin{equation}
\frac{d X_i}{dt} = G_i({\vec X}) - X_i H_i({\vec X})
\label{gen}\end{equation}
where the functions $G$ and $H$ are functions in principle of all the $X_i$ variables.
As a simple example, consider the $SIR$ model of Eqs.~(\ref{sirmf}):
\begin{equation}
\begin{array}{ll}
\displaystyle \frac{dS}{dt} -\beta SI & G_S=0; \ H_S=\beta I \\
 & \\
\displaystyle \frac{dI}{dt} = \beta SI -\gamma I & G_I=\beta SI; \ H_I=\gamma \\
& \\
\displaystyle\frac{dR}{dt} = \gamma I  & G_R=\gamma I; \ H_R=0
\end{array}
\end{equation}
The {\it JSON} files used specify the $G_i$ and $H_i$ functions for each state variable. Example
{\it JSON} files for the $SIR$ mean field and pair approximation equations follow:
\subsubsection{$SIR$ mean field equations}
The mean field equations of Eqs.~(\ref{sirmf}) for the $SIR$ model are described by the following
{\it JSON} file.
\begin{verbatim}
##############################################
# file sir_mf.json, derived from sir.json.
{
   "I" : {
      "G" : "beta*[S]*[I];",
      "H" : "gamma;"
   },
   "R" : {
      "G" : "gamma*[I];"
   },
   "S" : {
      "H" : "beta*[I];"
   },
   "first" : "S",
   "parameters" : [
      "beta",
      "gamma"
   ]
}
##############################################
\end{verbatim}
Some items to note about this:
\begin{itemize}
\item
Except for the {\tt "parameters"} and {\tt "first"} keys, which will be explained
shortly, the top-level keys of this file are the names of the singlets in
the model.
\item
The names of the singlets are the top--level keys of this file.
The corresponding value is a
container containing at least one, and possibly both,
{\tt "G"} and {\tt "H"} keys, which define the equation associated with that
singlet. Singlets appearing in the $G$ and $H$ functions must be enclosed within
square brackets. Remember that the factor of {\tt X} must be factored out of 
the {\tt H} equation for the {\tt X} singlet. All equations must be valid 
Matlab/Octave code, including the ending semicolons. If it is desired to split
up an equation over multiple lines for ease of readability, the 
associated value of the {\tt "G"} or {\tt "H"|} key is a list containing
the desired lines, as in, for example, this snippet:
\begin{verbatim}
   "I_T" : {
      "G" : [
         "p*beta*[S]*[I_U] + ...",
         "p*beta*delta_T*[S]*[I_T];"
      ],
      "H" : "mu_T + alpha_T;"
   },
\end{verbatim}
\item
The redundant equation (associated usually with the equation of $R$
in the $SIR$ model), which in principle is derivable from the fact
that the population remains constant, must be included in the {\it JSON} file.
\item
Parameters used in the equations must be declared through a
{\tt "parameters"} key, whose value is a list of the parameters
appearing in the equations.
\item 
The purpose of the {\tt "first"} key arises because key/value pairs
in a container will not be sorted. If a {\tt first"} key is present
in this {\it JSON} file, the singlets in the Matlab/Octave
update functions to be ordered such that the value of the {\tt "first"} key
appears first, after which the rest are sorted alphabetically. If a
{\tt "first"} key is not present, the singlets will be sorted alphabetically.
\end{itemize}
\subsubsection{$SIR$ pair approximation equations}
A {\it JSON} file describing Eqs.~(\ref{sirpa}) of the $SIR$ model in the pair approximation
is as follows:
\begin{verbatim}
##############################################
# file sir_pa.json, derived from sir.json.
{
   "II" : {
      "G" : "2*tau*([ISI] + [SI]);",
      "H" : "2*gamma;"
   },
   "IR" : {
      "G" : "tau*[RSI] + gamma*[II];",
      "H" : "gamma;"
   },
   "RR" : {
      "G" : "2*gamma*[IR];"
   },
   "SI" : {
      "G" : "tau*[SSI];",
      "H" : "tau*([ISI] + 1) + gamma;"
   },
   "SR" : {
      "G" : "gamma*[SI];",
      "H" : "tau*[RSI];"
   },
   "SS" : {
      "H" : "2*tau*[SSI];"
   },
   "first" : "S",
   "pa_parameters" : {
      "beta" : "tau"
   },
   "parameters" : [
      "tau",
      "gamma"
   ],
   "singlets" : [
      "S",
      "I",
      "R"
   ]
}
##############################################
\end{verbatim}
The comments made about the mean field {\it JSON} file of the previous section also apply here. In 
addition, note that
\begin{itemize}
\item
Except for the {\tt "first"}, {\tt "pa\_parameters"}, {\tt "parameters"}, and
{\tt "singlets"} keys, the top-level keys of this file are the
names of the doublets.
\item
Doublets appearing in {\tt G} and {\tt H} are denoted by {\tt [SS]}, {\tt [SI]}, 
etc. 
\item
Triplets appearing in {\tt G} and {\tt H} are denoted by {\tt [SSI]}, {\tt [RSI]}, 
etc. In the Matlab/Octave code these 
will be translated into appropriate function calls implementing
the Keeling approximation of Eq.(\ref{keeling}) for triplets in terms of doublets and
singlets.
\item
The meaning of the {\tt "parameters"} and the {\tt "first"} keys are equivalent
to that used for the singlet case.
\item
Items specified by a {\tt "pa\_parameters"} key, which are of the
form {\tt s1: d1}, {\tt s2: d2}, etc. will
specify a set of parameter pairs {\tt s1, d1}, {\tt s2, d2}, etc, where 
{\tt s1} is a parameter which would appear in the mean field model
and {\tt d1} is a derived parameter {\tt d1 = s1 / n} appearing in the
pair approximation model. This line is optional, and is only used
in providing an example main file.
\item
The list corresponding to the {\tt "singlets"} key is the list of
singlet keys.
\end{itemize}
\subsubsection{The {\tt maple\_chk.pl} utility program}
Once the {\it JSON} file describing the equations is generated, either for the
mean field or pair approximation equations, it may be useful at this
stage to run the supplied utility Perl script {\tt maple\_chk.pl} on it.  This addresses
two points present in the {\it JSON} file:
\begin{itemize}
\item
The values of the {\tt "G"} and {\tt "H"} keys the {\it JSON} file are valid Matlab/Octave
code, but especially if they are generated by {\tt make\_equns.pl}, there
may be some algebraic simplifications possible that are not implemented.
\item
By including the normally ``redundant'' state variable, there is a combination
of the equations that vanishes identically, reflecting
the fact that the population remains constant when
all possible transitions are taken into account. For example,
in Eqns.~(\ref{sirmf}) for the $SIR$ mean field model, we have the
combination of terms:
\begin{equation}
\frac{dS}{dt} + \frac{dI}{dt} +\frac{dR}{dt}
=  (-\beta SI) + (\beta SI -\gamma I) + (\gamma I ) = 0
\end{equation}
whereas for Eqns.~(\ref{sirpa}) for the $SIR$ model in the pair approximation, this is
\begin{equation}
\frac{d\, [SS]}{dt} +
2\frac{d\, [SI]}{dt} +
2\frac{d\, [SR]}{dt} + 
\frac{d\, [II]}{dt} +
2\frac{d\, [IR]}{dt} +
\frac{d\, [RR]}{dt} =0
\end{equation}
which uses the property of $\sum_B [AB] = n[A]$ from Eqns.(\ref{sum}), as well as
$[AB] = [BA]$. Note that
the assumption of Eq.~(\ref{keeling}) relating the triplets to doublets and singlets is
not used here. 
\end{itemize}
If one uses the {\it JSON} file describing the equations as input into the {\tt maple\_chk.pl}
Perl script, via running, for example,  {\tt perl maple\_chk.pl sir\_mf.json} for the
mean field equations of the SIR model, a text file {\tt  sir\_mf\_maple.txt} is
generated. For this $SIR$ mean field equations, this file is
\begin{verbatim}
 VAR_I_G := beta* VAR_S * VAR_I :
 VAR_I_H := gamma:
 VAR_R_G := gamma* VAR_I :
 VAR_S_H := beta* VAR_I :
 VAR_I_G := simplify( VAR_I_G );
 VAR_I_H := simplify( VAR_I_H );
 VAR_I_prime :=  simplify(VAR_I_G - VAR_I * ( VAR_I_H ));
 VAR_R_G := simplify( VAR_R_G );
 VAR_R_prime :=  simplify(VAR_R_G);
 VAR_S_H := simplify( VAR_S_H );
 VAR_S_prime :=  simplify( - VAR_S * ( VAR_S_H ));
 SUMALL := simplify(VAR_I_prime + VAR_R_prime + VAR_S_prime);
\end{verbatim}
This text file can be loaded as {\it "Maple Input"} into Maple 
and evaluated. Some features to note about this file are:
\begin{itemize}
\item
All variables, except for the last {\tt SUMALL} variable, have a {\tt VAR\_}
prefix. The Maple variable {\tt VAR\_I\_G}, for example, represents the
{\tt "G"} term for the {\tt "I"} node, {\tt VAR\_R\_H} represents the {\tt "H"}
term for the {\tt "R"} node, and so on.
\item
A series of Maple variables with a {\tt \_prime} suffix are introduced,
representing the equation for the specified node. For example,
{\tt VAR\_I\_prime} in the Maple code
\begin{verbatim}
 VAR_I_G := beta* VAR_S * VAR_I :
 VAR_I_H := gamma:
 VAR_I_prime :=  VAR_I_G - VAR_I * ( VAR_I_H );
\end{verbatim}
represents the equation
\begin{equation}
\frac{dI}{dt} = \beta SI -\gamma I\
\end{equation}
\item
The Maple variable {\tt SUMALL := VAR\_I\_prime + VAR\_R\_prime + VAR\_S\_prime},
which is always the last variable defined in the text file, represents
the sum of all of the equations
\begin{equation}
\frac{dS}{dt} + \frac{dI}{dt} +\frac{dR}{dt}
\end{equation}
which normally vanishes algebraically.
\item
When the worksheet is evaluated, the presence of the {\tt simplify} 
Maple function will cause all variables to be displayed in their
simplified form. 
\end{itemize}
\subsubsection{Running {\tt make\_mfile.pl} }
Having constructed the {\it JSON} file describing the equations, the
{\tt make\_mfile.pl} script is then run to generate Matlab/Octave functions
that can be used to solve the equations. The script may be used as follows; in these
examples, the input {\it JSON} file is denoted {\tt sir\_eqns.json}.
\begin{itemize}
\item
{\tt perl make\_mfile.pl sir\_eqns.json}: 
This will generate two files: {\tt sir\_eqns.m}, a Matlab/Octave file for the model, 
and {\tt sir\_eqns\_main.m},
an example main program.
\item 
{\tt perl make\_mfile.pl --input sir\_eqns.json --mfile mfile.m --main main.m}:
This does the same as the previous usage, but allows you to specify the
output names.
\item
{\tt perl make\_mfile.pl --help}:
This gives a brief summary of usage.
\item
{\tt perl make\_mfile.pl --gen}:
This generates a Matlab/Octave file {\tt de\_solve.m}, which is a file containing
functions needed to solve the equations, in addition to the output {\tt sir\_eqns.m}
file. The {\tt de\_solve.m} file is independent of the model under consideration,
and thus need be generated only once.
\end{itemize}
\subsubsection{Update algorithm }
The procedure used to solve the system of equations represented by Eq.(\ref{gen}) is
coded in the output Matlab/Octave file when running this script. The algorithm is as
follows. The continuous time variable is approximated by a series of discrete
steps $t_k$, with $k=1, 2, \ldots, K$. At any given time step $k$, the $i^{\rm th}$  equation 
of the $i=1, 2, \ldots, N_X$ set is
then approximated as
\begin{equation}
\frac{dX_i}{dt} \approx \frac{ X_i^{k+1} - X_i^k}{\Delta t} = 
G({\vec X_{\rm up}} ) 
-X_i^{k+1} H({\vec X_{\rm up}} )
\label{genapp}\end{equation}
where $\Delta t = t_{k+1} - t_k$, which is assumed constant and positive, and 
\begin{equation}
{\vec X_{\rm up}} \equiv (X_1^{k+1}, X_2^{k+1}, \ldots, X_i^{k} X_{i+1}^k\ldots X_{N_X}^k)
\end{equation}
In this, we have assumed that the update routine has already been performed on the
previous equations for $X_1, X_2, \ldots, X_{i-1}$. Eq.(\ref{genapp}) then allows
us to solve for the unknown variable $X_i^{k+1}$:
\begin{equation}
X_i^{k+1} = \frac{X_i^k + \Delta t\  G({\vec X_{\rm up}} ) }
{ 1 + \Delta t\  H({\vec X_{\rm up}} ) }
\end{equation}
for $k=1, 2, \ldots, K$, with the first $k=1$ value specified by the initial conditions. Since $G$ and
$H$ are ono--negative functions, Eq.(\ref{genapp}) ensures that
$X_i^{k+1}$ is positive semi-definite if $X_i^{k}$ is.
\subsubsection{Representation of singlets and doublets}
Within the Matlab/Octave code, singlets and doublets are represented as arrays {\tt X(i)}. 
The array index $i$ is defined as follows:
\begin{itemize}
\item
For a system with $N_S$ singlets, an array index $i=1, 2, \ldots, N_S$ is assigned according to
the order that the singlets are defined within the input {\it JSON} file, which is described under the
option of the {\tt first} key.
\item
For a system with $N_K$ doublets, with $N_K =  N_S(N_S+1)/2$ for $N_S$ singlets, indices are
assigned as follows. A singlet index is first generated based on the order given in the input
{\it JSON} file. Then, for a doublet $[S_iS_j]$ corresponding to singlets $S_i$ and $S_j$, an index
$1 \le P(i, j) \le N_K$ is generated:
\begin{equation}
P(i, j) =  \frac{1}{2}(2N_S-L)(L-1) + L + | i-j |
\end{equation}
where $L\equiv {\rm min}(i, j)$. Note that $P(i, j)$ is symmetric in $i$ and $j$, so that the doublet
$[S_iS_j]$ is mapped to the same index as the doublet $[S_jS_i]$, in accordance with the
relation $[AB] = [BA]$ of Eq.~(\ref{rels}). Making the choice $i \le j$,
the order of the assigned indices for the doublets is thus
determined by the order of the assigned indices for the singlets.
\end{itemize}
For the triplets appearing in the equations for the pair approximation,
using the property $[ABC] = [CBA]$ of
Eqs.~(\ref{rels}), a linear addressing scheme for a triplet $[S_iS_jS_k]$ is
\begin{equation}
T(i, j, k) = (j-1) N_K +P(i, k)
\end{equation}
which we write in abbreviated form as
\begin{equation}
(ijk) = (j-1)N_K + (ik)
\end{equation}
where $(ijk) \equiv T(i, j, k)$ and $(ij) \equiv P(i, j)$. 
Triplets appearing in the equations can then be expressed in terms of singlets and doublets using
the Keeling approximation of Eq.(\ref{keeling}). If we denote the singlets by $y_i$, the 
doublets by $x_{(ij)}$, and the triplets by $z_{(ijk)}$, then we have
\begin{eqnarray}
y_i &=& \frac{1}{n} \sum_{j=1}^{N_K} x_{ (ij) } \nonumber\\
z_{ (ijk) } &=& 
\zeta \frac{ x_{ (ij) }  x_{ (jk) } } { y_j } \left[ (1-\phi) + 
\phi \frac{N}{n} \frac{ x_{ (ik) } } { y_i y_k} \right]
\end{eqnarray}
In this way the triplets can be
related to the singlets and doublets.
In the update algorithm used in the Matlab/Octave functions generated by
{\tt make\_mfile.pl}, if a triplet appears in the 
{\tt G} term, a call to a function {\tt ZZ(...)} is made,
while another function {\tt ZZ1(...)} is used if it appears in the {\tt H} term. The difference between these
two functions is that {\tt ZZ1(...)} factors out the necessary {\tt X(i)} factor, while 
{\tt ZZ(...)} does not.
\subsubsection{Main program}
As well as generating the Matlab/Octave file containing the necessary functions to solve the
equations, {\tt make\_mfile.pl} also generates a sample main program. It is in this program
that the parameter values and initial conditions are specified, as well as calls made to
the functions needed to solve the equations. For a given system of equations, after running
{\tt make\_mfile.pl} on the {\it JSON} file, the only file that needs editing will be the sample main program.
\par
As an example, the following, with some comments removed,
 is the sample main program generated for the $SIR$ mean field
model.
\begin{verbatim}
%%%%%%%%%%%%%%%%%%%%%%%%%%%%%%%%%%%%
%  The following lines must be edited
%
% You may want to uncomment the following line to clear memory
% clear all
%
%  Give the values of the parameters of the equations:
beta = 0.0;
gamma = 0.0;
% Give the initial conditions:
X0.S = 0.0;
X0.I = 0.0;
X0.R = 0.0;
% specify the range of time desired for the solution:
%    t0 = initial time, t1 = final time, numpts = number of points between
t0 = 0.0;
t1 = 0.0;
numpts = 0.0;
tspan = linspace(t0, t1, numpts);
%
% end of user input
%%%%
% The following lines should not need editing
%
% put the parameter values in a structure data
%
data.beta = beta;
data.gamma = gamma;
%%%%
% call the routine to solve the equations
[t, X, chk] = de_solve(@sir_mf, data, tspan, X0);
% plot the first two structure members of X, plus the check
figure;
plot(t, [X.S], t, [X.I], t, chk);
%%%%%%%%%%%%%%%%%%%%%%%%%%%%%%%%%%%%
\end{verbatim}
Some features illustrated here are as follows.
\begin{itemize}
\item 
The values of the parameters specified in the {\tt parameters} key of the {\it JSON} file
are set here.
\item 
In the main program, the state variables $X_i$ are represented as a structure. For the
mean field case, the keys of the structures are the name of the singlets, while for the
pair approximation equations, the keys are the names of the doublets. For faster
execution speed, such structures are converted within the Matlab/Octave functions
to arrays with integer indices. The input initial conditions are specified by a
structure, say, {\tt X0.key}, while the output variable {\tt X.key(i)} is a structure with
key corresponding to the state variable and index $i$ corresponding to the time step.
\item
The initial and final times are specified by the variables {\tt t0} and {\tt t1}. An equally
spaced array of time steps {\tt tspan(i)}, with {\tt i=1, 2, ..., numpts} is then generated
by a call to the {\tt linspace} function. If {\tt tspan} consists of an array with 2 members
(the initial and final time), an array is generated within the Matlab/Octave code
with {\tt numpts} set to a specified value, to be described shortly.
\end{itemize}
The routine to solve the mean field equations is called as
\begin{verbatim}
 [t, X, chk] = de_solve(@sir_mf, data, tspan, X0);
\end{verbatim}
Recall that {\tt de\_solve.m} is generated by running {\tt perl make\_mfile.pl --gen}.
The input arguments are
\begin{itemize}
\item {\tt @sir\_mf}, which is a function handle pointing to the {\tt sir\_mf.m} file generated
when running {\tt make\_mfile.pl};
\item {\tt data}, which is a structure with keys corresponding to the parameters;
\item {\tt tspan}, which is an array specifying the desired range of time steps;
\item {\tt X0}, which is the structure {\tt X0.key} described above specifying the initial conditions.
\end{itemize}
The output variables, which for the mean field case, are {\tt  [t, X, chk]}. These are
\begin{itemize}
\item {\tt t}, which is an array {\tt t(i)} containing the range of times.
\item {\tt X}, which is the structure of arrays {\tt X.key(i)} giving the solutions to the equations
at the specified time steps.
\item {\tt chk}, which is an array {\tt chk(i)} containing the sum of all state variables at each time
{\tt t(i)}. As normally this is a set constant, equal to the number of nodes in the network, this
provides a useful check on the numerical accuracy.
\end{itemize}
A main program for the case of the equations in the pair approximation has similar structure.
One difference between the mean field and pair approximation cases is that, in the pair
approximation, {\tt de\_solve} is called as
\begin{verbatim}
 [t, X, Y, chk] = de_solve(@sir_pa, data, tspan, X0);
\end{verbatim}
The additional output variable {\tt Y} in this case is a structure {\tt Y.key(i)}, which is the
singlet calculated from the doublets according to the second equation of Eqs.(\ref{rels}).
\par
The {\tt de\_solve} function can accept, for either the
mean field or pair approximation case,  an optional 5th argument of a
structure, such as in 
\begin{verbatim}
[t, X, Y, chk] = de_solve(@sir_pa, data, tspan, X0, opts);
\end{verbatim}
This structure can be used to control some of the parameters involved in
solving the system of differential equations. The recognized members of the
structure are
\begin{itemize}
\item
{\tt opts.numpts}:
This option, which has a default of 100, specifies the number of points to use
if the {\tt tspan} array only contains two points (ie, the desired start and end times).
\item
{\tt opts.adaptive}:
This option, which is boolean and has a default of {\tt false}, specifies that an
adaptive routine should be used that adjusts the step size when 
solving the equations.
In difficult cases, setting this option to {\tt true} may improve the 
accuracy, but at
the expense of a speed penalty and larger memory usage.
\item
{\tt opts.maxerr}:
This option, which has a default of 0.001, is used when 
{\tt opts.adaptive} is {\tt true} to specify the desired error level.
\item
{\tt opts.maxit}:
This option, which has a default of 40, is used when {\tt opts.adaptive}
is {\tt true} to specify the maximum number of iterations tried 
when attempting to reach the desired error tolerance in a given interval.
\end{itemize}
\section{Conclusions}
The authors would be happy to receive correspondence regarding these programs, 
including bug reports and suggestions. The code in these programs is free software; 
you may redistribute and/or modify it 
under the same terms as Perl itself. See
\url{http://www.opensource.org/licenses/artistic-license-2.0.php} for details.
\section*{Acknowledgements}
This work was supported by the Natural Sciences and Engineering Research
Council of Canada.
\section*{Appendix}
The full {\it JSON} file for the drug resistant model described in Section \ref{drsection}
appears below.
\begin{verbatim}
##############################################
# file dr.json
{       
   "S": [
      {
         "target": "I_U",
         "link": { 
            "I_U": "(1-p)*beta", 
            "I_T": "(1-p)*beta*delta_T",
         },
         "needs": "I_U"
      },
      {
         "target": "I_T",
         "link": {
            "I_U": "p*beta",
            "I_T": "p*beta*delta_T",
         },
         "needs": "I_T",
      },
      {
         "target": "I_R",
         "link": "beta*delta_R",
         "needs": "I_R"
      },
   ],
   "I_U": {
       "target": "X",
       "link": "mu_U",
   },
   "I_T": [
      {
        "target": "X",
        "link": "mu_T",
      },
      {
        "target": "I_R",
        "link": "alpha_T",
      },
   ], 
   "I_R": {
      "target": "X",
      "link": "mu_R",
   },
   "pa_parameters": {
      "beta": "tau"
   },
   "first" : "S",
}
##############################################
\end{verbatim}

\end{document}